\documentclass[
  journal=mdy,
  manuscript=research-article,  
  year=2026,
  volume=x,
]{cup-journal}

\usepackage{amsmath}
\usepackage[nopatch]{microtype}
\usepackage{booktabs}
\usepackage{amsmath,amssymb,amsfonts}
\usepackage{algorithmic}
\usepackage{graphicx}
\usepackage{textcomp}
\usepackage{xcolor}
\usepackage{subcaption}
\usepackage{url}
\usepackage{bm}
\usepackage{graphicx}
\usepackage{subcaption} 
\usepackage{longtable}
\usepackage{hyperref}
\hypersetup{
	colorlinks=false,          
	citebordercolor=[rgb]{0, 0.0, 0.9}, 
	linkbordercolor=[rgb]{0.2, 0.8, 0.1}, 
	urlbordercolor=[rgb]{0, 0.7, 0.6},  
	pdfborder={0 0 1}          
}

\newcommand{\eurusd}{\mathtt{eurusd}}
\newcommand{\usdjpy}{\mathtt{usdjpy}}
\newcommand{\audjpy}{\mathtt{audjpy}}
\newcommand{\eurjpy}{\mathtt{eurjpy}}

\title{Macroeconomic Message Passing for Anticipating Foreign Exchange Regime Changes: A Deep Logical Learning Approach using Graph Tsetlin Machines}

\author{Christian Blakely}
\affiliation{Centre for Artificial Intelligence Research, University of Agder, Norway}
\email[Christian Blakely]{blakely@bernoly.com}
\alsoaffiliation{Head of AI, Bernoly AG, Zurich, Switzerland} 

\author{Melanie Gilmore}
\affiliation{La Jolla Private Wealth Group, Wells Fargo Advisors, La Jolla, CA, USA}


\begin{document}

\begin{abstract}
This paper introduces a graph-theoretic approach for predicting market regimes in foreign exchange (FX) currency prices. Specifically, the proposed model incorporates exogenous macroeconomic variables to update localized node features via message-passing operations. Utilizing the Graph Tsetlin Machine (GraphTM) framework, we empirically demonstrate the efficacy of this approach in anticipating market regimes for the US Dollar and Japanese Yen currency pair (USD/JPY). By representing multivariate macroeconomic drivers and technical indicators as hypervectorized directed multigraphs, the GraphTM leverages structured message passing to construct deep, interpretable logical clauses capable of recognizing complex sub-graph patterns.
\end{abstract}

\section{Introduction}\label{sec:introduction}

The identification of foreign exchange (FX) market regimes, namely distinct states of price action governed by specific macroeconomic conditions, is a challenging area of quantitative finance. Their prediction and estimation goes as far back to the 1980s with classical parametric models (\cite{Hamilton1989}; \cite{Krolzig1997}). They successfully introduced latent state tracking, but are structurally bound by strict linearity assumptions and scale poorly in high-dimensional feature spaces. Modern machine learning implementations such as the ones found in (\cite{TwoSigma2021}; \cite{erdogdu}) mitigate this by utilizing unsupervised learning over macroeconomic periods labeled by four different regime types. Similarly, Bank of England produced \cite{lee2023foreign} that evaluates FX forward hedging strategies for the Pound Sterling ($GBP$) across four distinct market states using an advanced regime-switching framework. However, all these papers tackled much longer time horizons and thus are much slower to adapt to quickly changing shocks in the regime over time. In this paper, we present a methodology for anticipating four different regimes that operate on hourly data. We will focus on the $\usdjpy{}$ bid/ask prices and demonstrate that our model can adapt to new information in the market quickly.

Recent advancements in Tsetlin Machines (TM) have introduced interpretable, logic-based pattern recognition that competes with deep learning architectures while maintaining high computational efficiency. However, the standard TM is constrained by a boolean representation of input data and fixed-length architectures with no sense of memory to the past, unless explicitly used as input without learning when it should or shouldn't be used and its importance or relevance (also known as ``attention"). To compensate for this setback, the Convolutional Tsetlin Machine (ConvTM) has been considered in market microstructure to do both regime detection in orderbooks in \cite{blakely2022tsetlin} as well as fast higher-order the microprice estimates in \cite{blakely2024highresolutionmicropriceestimates}.

The Graph Tsetlin Machine (GraphTM) introduced in \cite{granmo2026tsetlinmachinegoesdeep} made an impact in the logic-based machine learning community by offering a new architecture which processes multimodal data represented as hypervectorized directed multigraphs. This allows for a wide range of new applications for learning from data as there is no more constraint on the relationships between different types of data.

In this paper, we leverage this computational relationship structure by mapping market indicators to nodes and their causal and correlation relationships to edges. We build a GraphTM that allows context to be gathered over the market's topography rather than just temporal locality. This is achieved through a message-passing mechanism that constructs nested deep clauses, enabling the recognition of complex sub-graph patterns with significantly fewer logical rules.

\subsection{The Challenge of Interest Rate-Driven Stagnation}

A primary challenge in FX regime anticipation is the occurrence of prolonged periods of low variance, often driven by static interest rate policies. A notable example is the 2018–2019 period, where a lack of divergence in central bank policies led to a significant "volatility drought" in the $\usdjpy{}$ pair. During such intervals, traditional signal-to-noise ratio (SNR) metrics become compressed. Anticipating improvements in SNR conditions associated with monetary policy changes or inflation surprises may enhance risk management and portfolio allocation decisions. 

Due to the sensitive nature of interest rate divergence on the $\usdjpy{}$, we will focus on this currency pair throughout this study. As such, we propose that the GraphTM can be an attractive alternative to traditional machine learning models for regime detection due to two key technical characteristics of the GraphTM architecture:
\begin{enumerate}
    \item \textbf{Reasoning by elimination:} Through its layer-zero clause components, the GraphTM can specify the absence of specific properties (e.g., absence of volatility expansion) to accurately classify stagnant states using fewer clauses.
    \item \textbf{Noise robustness via hypervectors:} By encoding node properties and messages as symbols in a high-dimensional hypervector space, the GraphTM maintains robust predictive capabilities even in environments with extremely low signal clarity.
\end{enumerate}

With this in mind, we demonstrate that by incorporating temporal trajectories of deltas (changes in price characteristics) into the graph structure, the GraphTM can distinguish between a true stagnant regime ahead and the initial "acceleration" of an incoming trend, even when absolute volatility levels remain historically low.

As we will see, the GraphTM architecture allows for the integration of temporal trajectories and cross-asset correlations through nested logical rules. We develop a systematic detection system based on a 72-hour majority-vote regime labeling approach to evaluate the model's performance. A 72-hour horizon was selected to balance two competing objectives. Shorter horizons frequently produced unstable regime assignments that were highly sensitive to temporary price fluctuations, whereas substantially longer horizons reduced the responsiveness of the labeling framework to evolving macroeconomic conditions. Preliminary exploratory analysis indicated that a three-day window provided a reasonable compromise between the temporal stability and responsiveness.

Our results suggest that the GraphTM's ability to filter out stochastic noise through reasoning by elimination significantly improves regime classification accuracy over traditional approaches. Specifically, the model suggests superior resilience during the characteristic "volatility droughts" of 2018–2019 and improved anticipation of the high-volatility trending periods post-2021. The objective of this work is to explore the applicability of GraphTMs to FX regime anticipation and to evaluate whether graph-based logical learning can provide an interpretable alternative to conventional machine learning approaches. More broadly, the study investigates how macroeconomic relationships, represented explicitly as graph structures, can be incorporated into symbolic learning systems for financial forecasting tasks. 

The remainder of this paper is organized as follows. Section \ref{sec:graph} introduces the architecture of the Graph Tsetlin Machine (GraphTM) and the learning algorithm, providing formal definitions for the node topography and directed edge sets that constitute our market graph. We also review the the feedback mechanisms and the construction of deep conjunctive clauses.  Section \ref{sec:labeling} discusses our labeling strategy, and how we define the regimes that we will learn to anticipate. We discuss not only the quantitative definitions of each regime, but also the qualitative features that one can visualize and recognize when they occur. Finally, Section \ref{sec:numerical_results} presents some numerical studies to demonstrate the empirical performance of the model, where we also compare it with benchmarks and many other learning paradigms such as Hidden Markov Models and Deep Learning. 

In \ref{sec:features}, more details on the construction of the features for the nodes which are used in the graph structure are given. \ref{sec:algorithm} reviews the GraphTM algorithm for updating and learning at each level. As our main computational structure in the GraphTM, we review the sparse binary hypervector framework in \ref{sec:hypervectors}, explaining the symbolic algebra and embedding strategies used to encode continuous market data and categorical relationships into a high-dimensional space in the nodes and edges.

\section{Functional Components of the GraphTM}\label{sec:graph}

The predictive mechanism of the vanilla TM is based on the construction of conjunctive clauses that identify patterns within a propositional feature vector $X = [x_1, x_2, \ldots, x_o]$. A clause is represented as a conjunction of literals, where each literal corresponds either to the presence or absence of a feature. It has the form $C_j = \bigwedge_{l_k \in L_j} l_k$, where $L_j$ represents a subset of literals consisting of features and their negations: $L_j \subseteq \{x_1, \ldots, x_o, \lnot x_1, \lnot x_2, \ldots, \lnot x_o\}$. All literals in this architecture are elements of a sparse binary hypervector (see \ref{sec:hypervectors} for definitions of our hypervector system). Transitioning to the GraphTM, clauses shift their operation from global feature vectors to local node properties $\mathcal{P}$. A clause literal $p_k$ or $\lnot p_k$ specifies the presence or absence of a property within a specific node. Within the proposed regime classification framework, a clause such as $C_j = p_{\text{\tiny lv}} \land p_{\text{\tiny le}}$ dictates that a node must simultaneously possess properties representing low volatility ($lv$) and a low efficiency ratio ($le$) for the clause to evaluate as true. Such clauses provide localized logical descriptions of market states and serve as the building blocks for higher-level relational patterns. 

To capture hierarchical market relationships, a GraphTM clause is further subdivided into components to create a deep clause, with one component allocated per layer: $C_j = C_j^0 \land C_j^1 \land \dots \land C_j^{D-1}$, where $D$ represents the depth of the network. The layer-zero components ($C_j^0$) operate directly on the primary node properties $\mathcal{P}$. This architecture is particularly well suited for identifying stagnant regimes through ``reasoning by elimination''. For example, a clause component can be trained to recognize a volatility drought by enforcing the absence of expansionary properties (e.g., $C_1^0 = \lnot p_{\text{\tiny ra}} \land \lnot p_{\text{\tiny ry}}$), effectively matching a ``stagnant'' state by eliminating active market characteristics such as rising average-true-range, which we will use as a price variance indicator (ATR, see \ref{sec:features} for definition), $ra$ or rising yields $ry$.

The connection between these layers is maintained by a set of message symbols $\mathcal{M}$. Whenever a component $C_j^0$ evaluates as true at a node, it utilizes a dedicated message symbol $M_j^0 \in \mathcal{M}$ to signal this state to its neighbors governed by edge type and direction. Every node $v_q$ in our market graph maintains an inbox $I_q^d$ to store these incoming signals from layer $d$. Subsequent clause components $C_j^i$ ($i > 0$) then inspect this inbox for messages $M_j^{i-1}$ from the previous layer to determine if the local node's state is supported by its neighbors' states. In our model, this allows a price node $v_{\text{\tiny UJ}}$ to verify its local stagnant classification by checking if neighboring bond nodes have transmitted messages indicating a corresponding lack of yield movement.

The process of message submission occurs every time a clause component produces a message, at which point it is dispatched to the inboxes of all adjacent nodes according to the graph’s defined edges $E$. For instance, if the US Yield node matches a ``stable'' property at layer zero, it sends a message $M_{\text{\tiny s}}^0$ to the inbox of the price node $v_{\text{\tiny UJ}}$. Upon receipt, the price node's layer-one clause component can be evaluated by comparing the inbox contents with the required message symbols to determine a match. This parallel processing of nodes and clauses across layers allows the model to compute the truth value of a complete multi-layer clause $C_j$ with high computational efficiency.

Furthermore, to distinguish between different types of economic influence, the GraphTM supports multiple edge types $t^t_{qr} \in \mathcal{T}$. Messages passed along our graph are annotated with their specific edge type—such as ``causal'' for interest rate drivers or ``correlative'' for cross-pair volatility—providing additional context that is stored within the node's inbox. This enables the deep clauses to differentiate between a price move driven by fundamental policy shifts versus one driven by technical spillover from auxiliary currency pairs.

Learning within the GraphTM is achieved through the interaction of decentralized Tsetlin automata distributed across nodes and layers. Rather than adjusting continuous-valued parameters through gradient optimization, the model modifies clause structures through discrete automata state transitions driven by feedback from classification outcomes. Unlike standard Tsetlin Machines that evaluate global feature vectors statically, the GraphTM updates the states of decentralized Tsetlin Automata (TA) distributed across layers and nodes, driving structural alignment via local and propagated feedback loops. THe GraphTM learning procedure can be viewed as consisting of two components: 1) evaluating and updated clause states given their outputs and 2) how to reward and penalize the literals in the clauses to learn based on feedback. See the original GraphTM paper by \cite{granmo2026tsetlinmachinegoesdeep}. The full algorithm can be found in \ref{sec:algorithm}.

\subsection{Tsetlin Automata and Clause State Updates}

Each literal $p_k$ or $\lnot p_k$ in a layer component $C_j^d$ is a feature of the binary hypervector structure and is governed by a dedicated Tsetlin Automaton. A TA possesses $2N$ states, where states $1$ to $N$ dictate the exclusion of the literal (Action: \textit{Exclude}), and states $N+1$ to $2N$ dictate its inclusion (Action: \textit{Include}). Let $S_{j,k}^{d,q}$ represent the integer state of the TA corresponding to literal $k$ of clause component $j$ at layer $d$ for node $v_q$. 

During the training phase, an input graph $G$ at time $T$ is presented alongside its 72-hour majority-vote target label $y_T$. The final classification is determined by a majority vote of the active clauses. Let $v_{\text{\tiny UJ}}$ be the target price node. The total vote for a particular class $c$ is given by:
\begin{equation}
	V^c(v_{\text{\tiny UJ}}) = \sum_{j \in \mathcal{C}^c} C_j(v_{\text{\tiny UJ}})
\end{equation}
where $\mathcal{C}^c$ is the set of clauses assigned to class $c$, and $C_j(v_{\text{\tiny UJ}})$ is the evaluation of the complete deep conjunctive clause spanning from the local node properties to the inboxes of its economic neighbors.

\subsection{Feedback Mechanisms}

The states of the TAs are modified using two distinct stochastic feedback loops: Type I (Inclusion/Generalization) and Type II (Discrimination). The type of feedback delivered to a node depends on the alignment between the clause's predicted class and the actual target label $y_T$, as regulated by a user-defined threshold parameter $T_{\text{\tiny vote}}$.

\subsubsection{Type I Feedback (Generalization)}
Type I feedback is applied to clauses assigned to the correct class when the voting sum is below the threshold ($V^c < T_{\text{\tiny vote}}$), or stochastically to smooth out noise. It penalizes literals that evaluate to False when the clause is active, forcing the model to simplify its logical requirements. 

For a local node component $C_j^0(v_q)$ evaluating to True, Type I feedback modifies the TA states according to:
\begin{equation}
	S_{j,k}^{0,q} \leftarrow 
	\begin{cases} 
		\min(2N, S_{j,k}^{0,q} + 1) & \text{if literal } k \text{ is True with probability } \frac{s-1}{s} \\
		\max(1, S_{j,k}^{0,q} - 1) & \text{if literal } k \text{ is False with probability } \frac{1}{s}
	\end{cases}
\end{equation}
where $s$ is a specificity parameter controlling the legal granularity. Type I feedback promotes the inclusion of properties that contribute consistently to correct classifications while reducing the influence of transient unstable patterns. Within the context of regime classification, this encourages the learning of recurring structural characteristics associated with prolonged periods of low market activity. 

forces the model to retain basic properties—such as the absence of volatility expansion ($\lnot p_{\text{\tiny ra}}$)—while discarding transient micro-signals, thus ensuring stable generalization during prolonged periods of low variance.

\subsubsection{Type II Feedback (Discrimination)}
Type II feedback is applied to clauses when the model incorrectly predicts their corresponding class. It introduces strict discrimination by reinforcing literals that evaluate to False, quickly deactivating clauses that generate false positives in wrong regimes.

If a clause $C_j$ evaluates to True at node $v_{\text{\tiny UJ}}$ but the ground-truth target is a different regime, Type II feedback forces every excluded literal ($\le N$) that evaluates to False in the current market state to increase its state value toward inclusion:
\begin{equation}
	S_{j,k}^{d,q} \leftarrow \min(2N, S_{j,k}^{d,q} + 1) \quad \text{for all } k \text{ where } \text{Literal}_k = 0 \text{ and } S_{j,k}^{d,q} \le N
\end{equation}
This mechanism facilitates discrimination between regimes by encouraging the inclusion of literals that distinguish incorrectly activated clauses from the observed market state. Consequently, clauses associated with stagnant conditions become less likely to activate during emerging trend regimes. If a clause configured to identify a low-volatility state stays active during an incoming ``Directional Explosion'' (Class 3), Type II feedback immediately suppresses it by forcing the inclusion of literals that contradict the active trend features.

\subsection{Message-Passing Backpropagation of Rules}

An important characteristic of the GraphTM is that feedback may be propagated through graph relationships rather than through continuous gradient updates. As a result, corrective information can influence not only the target node but also neighboring nodes that contributed to the activation of higher-level clauses. When a layer-one component $C_j^1(v_{\text{\tiny UJ}})$ receives feedback based on the target price node's label, the error is not backpropagated via continuous gradients as in neural networks. Instead, the feedback is routed symbolically back to the neighbor nodes (e.g., $v_{\text{\tiny ry}}$) that originally populated the price node's inbox $I_{\text{\tiny UJ}}^1$.

If a deep clause requires a message $M_j^0$ from the bond yield node to validate a trend, but the trend fails to materialize, Type II feedback is routed directly to the TA team governing $C_j^0(v_{\text{\tiny ry}})$. This feedback modifies the automata governing the originating node, allowing the learned clause structure to better align transmitted messages with subsequent market outcomes observed during training.

\section{Graph Model and labeling}
\label{sec:labeling}
The proposed framework represents the state of the market at each observation time as a directed multigraph $G = (V, E, P, \mathcal{T})$, where nodes correspond to market variables, edges represent predefined relationships between variables, and node properties encode the corresponding market state both present and historically through various features and related securities. The $\usdjpy{}$ price node $v_{\text{\tiny UJ}}$ serves as the primary target node, aggregating information transmitted from both technical and macroeconomic factors. Given that we wish to classify the forthcoming regime over the next days given the current state in time, this effectively can be represented as a graph classification problem at each timestamp with new information. We will use several types of nodes in our graph that are defined as follows.

\begin{itemize}
    \item \textbf{Target Price Node ($v_{\text{\tiny UJ}}$):} Evaluates local log-returns through $C_j^0$ and integrates messages from drivers via $C_j^1$.
    \item \textbf{Volatility Trajectory Node ($v_{\text{\tiny ra}}$):} Encodes normalized Average True Range ATR and provides a measure of local volatility condition, allowing the model to distinguish between low- and high-variance environments.
    \item \textbf{Efficiency Trajectory Node ($v_{\text{\tiny le}}$):} Encodes the Efficiency Ration (ER) and provides information regarding the directional persistence of price movement relative to accumulated volatility.
    \item \textbf{Macroeconomic Driver Nodes ($v_{\text{\tiny ry}}$):} Represent US and Japanese bond yield deltas to capture fundamental policy divergence.
\end{itemize}

\begin{figure}  
    \centering
    \includegraphics[width=4in]{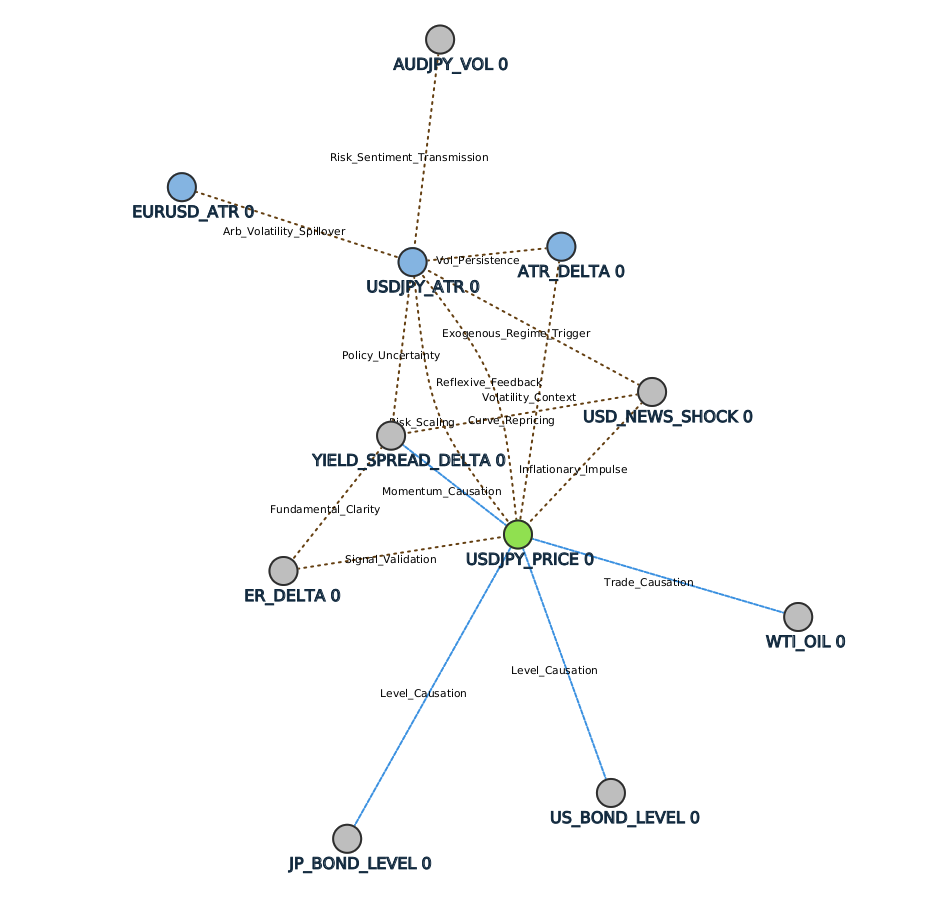}
    \caption{\footnotesize Illustrates the full graph topology adopted in this study, including the node types, input variables, and edge relationships used to facilitate information transmission across the network. We use normalized ATR and ER values of $\audjpy{}, \eurusd{}$ along with bond and Oil prices that can influence $\usdjpy{}$ supply and demand}
    \label{fig:graphUSDJPY}
\end{figure}

Figure \ref{fig:graphUSDJPY} shows the graph layout with the various state inputs at the nodes and the types of edges used.

Edges represent transmission channels. These allow for the construction of deep clauses that capture sub-graph patterns. These are the following edges we use along with their rationale. 
\begin{itemize}
    \item \textbf{Fundamental dependency ($v_{\text{\tiny ry}} \to v_{\text{\tiny UJ}}$):} Represents the assumed relationship between bond-yield dynamics and FX price movements. Can transmit messages signaling policy divergence.
    \item \textbf{Variance persistence ($v_{\text{\tiny ra}} \to v_{\text{\tiny UJ}}$):} Provides volatility context to filter out signals that lack requisite support.
    \item \textbf{Cross-Market volatility link ($v_{\text{\tiny CrossVol}} \to v_{\text{\tiny ra}}$):} Allows the model to learn how volatility in auxiliary pairs influence volatility in $\usdjpy{}$.
\end{itemize}

\begin{figure}  
	\centering
	\includegraphics[width=4in]{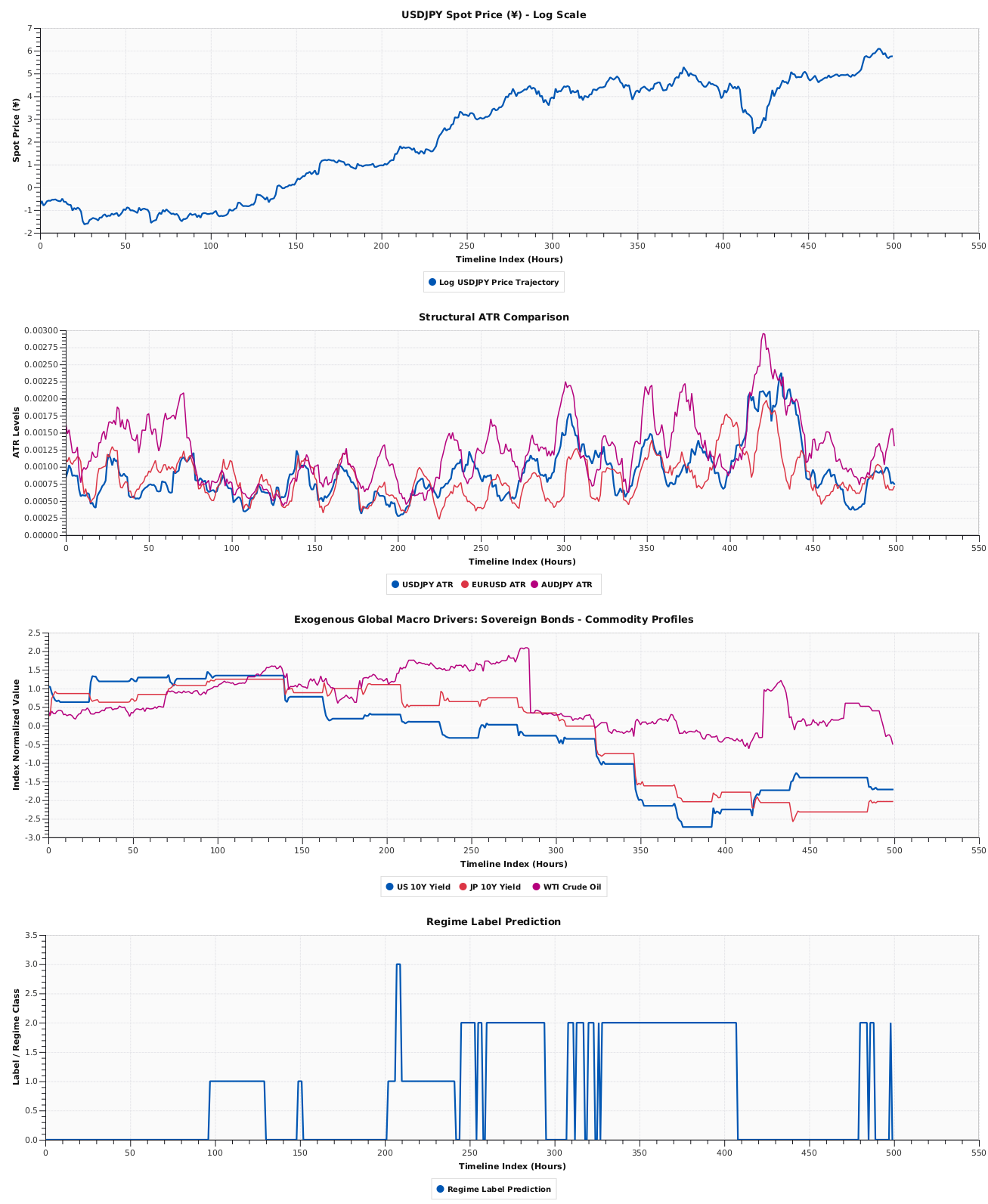}
	\caption{\footnotesize The different market data transformation on an hourly scale along 500 market trading hours. Starting from top to bottom, log price of the $\usdjpy{}$, the ATR features, the macro features, and the regime class prediction}
	\label{fig:dataplot}
\end{figure}

For the nodes $v_{\text{\tiny ra}}$, the normalized ATR and efficiency values will come from $\usdjpy{}$, $\audjpy{}$, and $\eurusd{}$, and are updated every hour. All volatility and efficiency-based measures are normalized prior to graph construction, reducing the need for additional in-sample scaling procedures. Our macroeconomic driver values, given by US and Japanese Bond prices along with Oil price, here we take the raw levels, where we have set the linear embedding into hypervector space using the ranges $[50, 120]$ with $40$ divisions for both, and $[40, 130]$ with $40$ divisions for crude oil prices. Additional details regarding feature construction and transformations are provided in  \ref{sec:features}. Figure \ref{fig:dataplot} shows the different market features used in the graph at the hourly time scale, along with the output of the predicted regime class for the next 72 hours. 

\begin{figure}[htbp]
	\centering
	\includegraphics[width=0.95\textwidth]{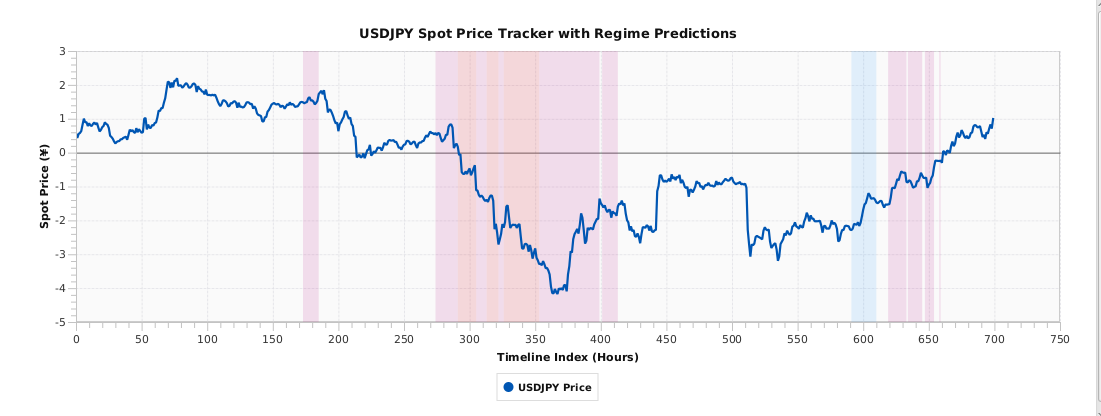}
	\caption{$\usdjpy{}$ spot price sequence paired with colored structural background regime tracking.}
	\label{fig:usdjpy_price_regimes}
\end{figure}

\begin{figure}[htbp]
	\centering
	\includegraphics[width=0.95\textwidth]{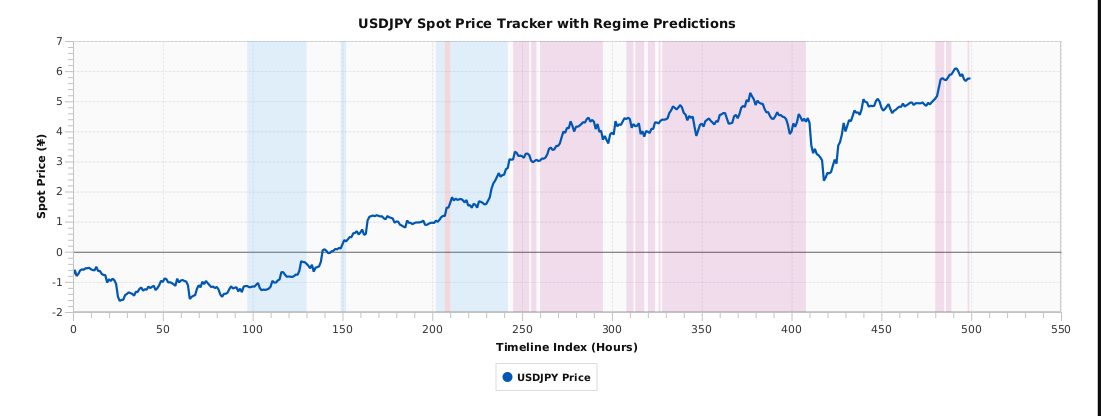}
	\caption{Changing from trending to mean-reversion prediction. Light blue represents low volatility trending.}
	\label{fig:multi_market_atr}
\end{figure}

\begin{figure}[htbp]
	\centering
	\includegraphics[width=0.95\textwidth]{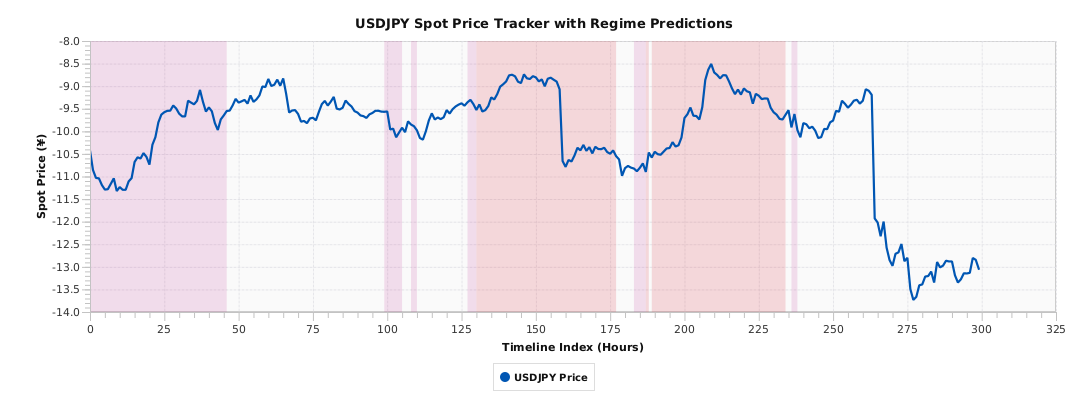}
	\caption{Exogenous Global Macro helping navigate between trending and high-vol mean reverting.}
	\label{fig:macro_drivers}
\end{figure}

For each new update in the data of the nodes in the graph, the state of the model potentially changes. The state is governed by the incoming market regime of the underlying ($\usdjpy{}$), which we define in a four-quadrant regime system. 

\begin{itemize}
    \item \textbf{Class 0: stagnant (low volatility / low efficiency):} This regime is characterized by narrow price ranges and high stochastic noise. Signal-to-noise conditions are generally weak on average, reflecting limited directional structure in price movements. The GraphTM utilizes reasoning by elimination to identify the absence of any growing or expansionary properties that could be coming from shifts in the market.
    
    \item \textbf{Class 1: steady trend (low volatility / high efficiency):} A state of consistent directional movement accompanied by low absolute variance. This often manifests during periods of stable interest rate differentials where carry trades can drive efficient but slow appreciation of a currency. These environments often exhibit higher directional efficiency relative to realized volatility. 
    
    \item \textbf{Class 2: choppy / mean-reverting (high volatility / low efficiency):} Characterized by large, frequent price swings that fail to produce a sustained directional trend. This regime may reduce the effectiveness of trend-following approaches.
    
    \item \textbf{Class 3: volatile trend (high volatility / high efficiency):} This regime represents high-conviction structural shifts and periods of elevated volatility with persistent directional movement, typically associated with major macroeconomic or monetary-policy adjustments.
    
\end{itemize}

\begin{figure}  
    \centering
    \includegraphics[width=4in]{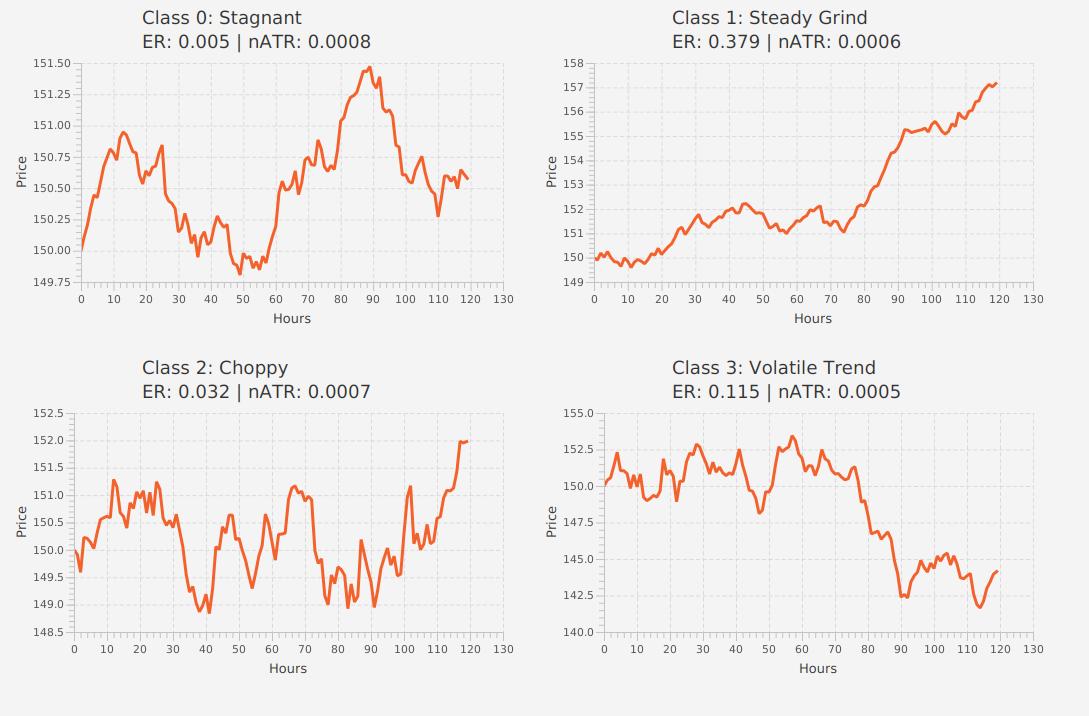}
    \caption{\footnotesize The four different regimes our model will learn to anticipate given current market conditions. All four regimes can be characterized by two features: normalized ATR and ER.}
    \label{fig:regimes}
\end{figure}

This taxonomy is constructed by intersecting two fundamental market characteristics: \textit{Hourly normalized volatility} and the \textit{Efficiency Ratio} (ER), which is a common SNR ratio. It is a key metric in our model for distinguishing between directional price moves and stochastic noise and measures the ratio between the net displacement of price and the total distance traveled over a specified window.  Figure \ref{fig:regimes} shows the four different regime classifications that our GraphTM model will learn to anticipate. 

Both regime measurements will be computed on an hourly time frame for both the ER and ATR nodes, as well as the look-ahead computation for the target over a 72-hour look-ahead window. To produce the labeling for our GraphTM model, we must characterize the state of the incoming regime given the current information available to all the nodes and edges in the graph. The objective of the learning process is to identify combinations of graph states that historically preceded sustained regime transitions. The instantaneous regime $\mathcal{R}(\tau)$ for any given hour $\tau$ is determined by the normalized ATR from the past 10 hours along with the ER measuring the net price change over a 72-hour window divided by the sum of absolute hourly changes; an $ER$ exceeding a threshold $\gamma$ signifies a trending state.

To ensure temporal stability and filter out transient hourly noise, the training label $y_T$ for a graph at time $T$ is defined by a majority vote over the subsequent 3-day window:
\begin{equation}
    y_T = \text{mode}(\{\mathcal{R}(\tau) \mid \tau \in [T+1, T+72]\})
\end{equation}
The use of a forward-looking majority-vote label encourages the model to associate current market conditions with the subsequent regime outcomes rather than contemporaneous price fluctuations. This construction seeks to reduce the influence of transient shocks and emphasize more presistent market transitions. For example, it can take several days before higher than expected CPI numbers contribute to a regime change, from stagnant (regime 0) to low volatility trending (regime 1). 

\begin{figure}  
    \centering
    \includegraphics[width=4.4in]{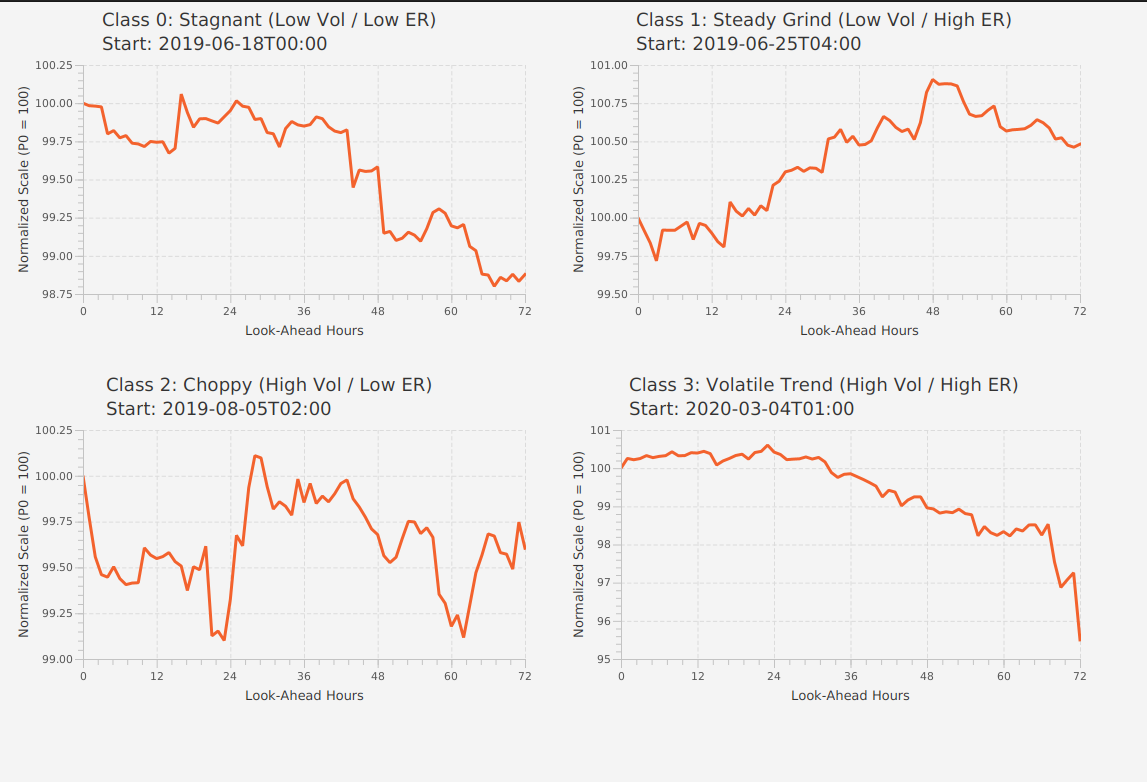}
    \caption{\footnotesize Four examples of regime labels on $\usdjpy{}$ over the 72 hour period used in the labeling. The date and hour of the beginning of the regime is shown for each plot. We used 100 JPY as the starting point to better show the magnitude of the movements.}
    \label{fig:examples}
\end{figure}

Figure \ref{fig:examples} show examples of the four different regime labels constructed by the 3-day look-ahead window for $y_T$. 

In class 0 (stagnant) we notice that the price only moves in a 25 pip range during the first two days and slow tapers off the third. In contrast, class 3 (high volatility and efficiency), the price trends down 500 pips consistently over 3 days, with a lot of big movements on day 3. In class 2, we see the choppy movements of high volatility at the hourly level with no big trends breaking out.

\section{Numerical Results and Empirical Evaluation}
\label{sec:numerical_results}

This section presents the empirical validation of the GraphTM framework applied to $\usdjpy{}$ regime anticipation. We detail the structured dataset partition, the chronological model training protocol, and the evaluation metrics utilized to gauge predictive efficacy. The primary objective of the empirical analysis is to evaluate the ability of the proposed framework to identify market regimes across varying volatility environments while maintaining out-of-sample predictive stability. Particular attention is given to periods characterized by prolonged low volatility and to intervals exhibiting structural regime transitions.

\subsection{Dataset Partitioning and Structural Isolation}

Evaluating predictive models on highly non-stationary financial time series requires strict adherence to chronological ordering to eliminate data leakage. Conventional random k-fold cross-validation may produce overly optimistic performance estimates for financial time series due to serial dependence and temporal overlap between observations.
To reduce the possibility of information leakage, we partition our continuous hourly historical dataset into two distinct, structurally isolated blocks of a training set (60\%) and an out-of-sample (OOS) test set (40\%). We obtained hourly data from all FX pairs, while taking pre-market, regular market, and after hours hourly data for all other features. All data were aligned on an hourly basis and any missing data for any feature is set to the previous recorded value. 

To maximize the model's adaptation to evolving macroeconomic structures while preserving strict OOS integrity, we deploy an anchored \textit{Walk-Forward Validation} strategy enhanced with data purging and embargo rules. The training process executes across rolling sliding windows. Because our target labeling mechanism uses a 72-hour look-ahead window ($T+72$), any training sample within 72 hours of the boundary between the training and validation sets contains future information spanning into the validation state. To eliminate this overlap, a 72-hour buffer of data points is completely deleted and purged at the trailing boundary of the training block.

To assess the sensitivity of model performance to hyperparameter selection, we conducted a 100-iteration random-search experiment spanning a broad range of clause counts, specificity values, and literal constraints. The model configuration landscape is sampled stochastically across predefined boundaries: total clause allocation is bound between $100 \le \mathcal{C} \le 400$, the max literal restriction is set to $20 \le \mathcal{L} \le 100$, and the clause specificity spans $2.0 \le s \le 20.0$. For simplicity and to maintain operational learning feedback, the voting threshold parameter $T$ is dynamically scaled as a random uniform fraction of the number of clauses.

For each generated configuration, the model is initialized chronologically on the training partition and validated against the OOS testing data. We aggregate the resulting classification performance vectors by calculating the distinct mean ($\mu$) and standard deviation ($\sigma$) of the prediction accuracies across all four targeted macroeconomic regimes. By tracking this dispersion, the resulting distribution of classification accuracies provides an indication of stability of model performance across different hyperparameter configurations.

\subsection{Quantifying exogenous macroeconomic feature performance gain }

\begin{figure}  
	\centering
	\includegraphics[width=3in]{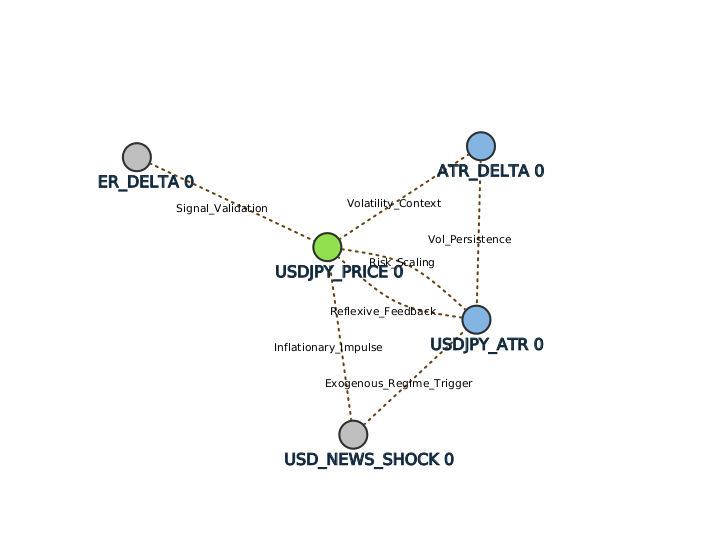}
	\caption{\footnotesize The reduced graph representation for testing feature importance of the additional macro market indicators.}
	\label{fig:reducedGraph}
\end{figure}

To evaluate the contribution of exogenous macroeconomic variables, we performed an ablation analysis comparing the full graph architecture with a restricted graph containing only local $\usdjpy{}$ features and direct news-related information, excluding any cross-market information.  The feature space for reduced model was strictly constrained to a 5-dimensional sub-space extracted exclusively from the target instrument $\usdjpy{}$ plus the direct fundamental news shock vector. This includes the historical ATR and ER values which measure the most recent trend buildup, if any. The reduced graph structure is shown in figure \ref{fig:reducedGraph} which includes nodes and edges only extracted from $\usdjpy{}$ hourly prices. 

All external nodes representing global macro liquidity (US and Japanese bond yield levels, complementary currency volatility profiles, and commodities ($\text{WTI\_Oil}$) were pruned from the graph network, alongside their respective edge mappings.

\subsection{Empirical Predictive Performance and Class Analysis}

The hyperparameter sensitivity simulation yielded a distinct performance profile across the four targeted macroeconomic regimes. Table~\ref{tab:random_search_results} outlines the out-of-sample prediction accuracies generated across the 100 random configurations on the testing partition.

\begin{table}[htbp]
\centering
\caption{GraphTM Out-of-Sample Accuracy Statistics (100-Iteration Random Search)}
\label{tab:random_search_results}
\begin{tabular}{lccc}
\hline
\textbf{Regime Class} & \textbf{Test Sample Count} & \textbf{Mean Accuracy ($\mu$)} & \textbf{Standard Deviation ($\sigma$)} \\ \hline
Class 0: Stagnant      & 33,052                      & 80.5\%                          & $\pm$ 3.3\%                            \\
Class 1: Steady trend  & 2,208                      & 48.1\%                          & $\pm$ 4.9\%                            \\
Class 2: Choppy        & 2,372                      & 71.9\%                          & $\pm$ 2.1\%                            \\
Class 3: Volatile trend& 385                        & 11.2\%                          & $\pm$ 1.6\%                            \\ \hline
\end{tabular}
\end{table}

Performance differed considerably across the four regime classes, reflecting both the underlying class imbalance and the varying difficulty associated with each forecasting task. The GraphTM achieves its highest predictive stability in class 0 (Stagnant, $\mu = 80.5\%$) and class 2 (Choppy, $\mu = 71.9\%$). The relatively small standard deviations associated with these classes could indicate that the decentralized conjunctive clauses converge onto stable, invariant representations of market noise and mean reversion. This suggests that properties such as compressed moving ranges ($v_{\text{\tiny ra}}$) and decaying efficiency trajectories ($v_{\text{\tiny ed}}$) present repetitive boolean states that the clauses of the nodes can exploit.

Conversely, class 1 (Steady trend) suggests a moderate but structurally viable predictive accuracy of $48.1\%$. The low dispersion across configurations highlights that the model maintains a consensus when tracking structural macro accumulation. Class 3 presents a bottleneck as it remains a statistical minority, accounting for only 385 OOS test records. The limited sample size, together with the abrupt onset and heterogeneous nature of high-volatility trend episodes, likely contributed to the reduced classification performance observed for this regime.

In order to determine the magnitude of the message passing GraphTM approach on regime detection, we now compare against four distinct algorithmic paradigms, including a Gradient Boosted model, sequential discrete Hidden Markov model, a graph neural network, and convolutional TMs (ConvTM) (see \cite{granmo2019convolutional}). We also compare it across dozens of models that are found in H2O AutoML (see \cite{ledell2020h2o}), which comprises of Boosted Gradient, DeepLearning, and other ML models typically used in supervised learning.  All models were trained on the same in-sample and then evaluated on the same out-of-sample (OOS) sequences using the leakage-protected sequential split framework. See Table \ref{tab:automl_massive_results} for the full version of the comparison table across all models. 

Since no major parameter hypertuning was done, we simply chose randomly 100 configurations for each model (number of trees, nodes, etc) and are reporting the best accuracy achieved out-of-sample. Code can be found on the first author's Github page linked below. The experimental outcomes across the four macroeconomic market regimes are compiled in Table~\ref{tab:benchmark_results}.

\begin{table}[htbp]
\centering
\caption{Comparative Out-of-Sample Performance Matrix Across Market Regimes}
\label{tab:benchmark_results}
\small
\begin{tabular}{lcccccc}
\toprule
\small
\textbf{Architecture} & \textbf{Overall Acc.} & \textbf{Class 0} & \textbf{Class 1} & \textbf{Class 2} & \textbf{Class 3} & \textbf{Interpretability} \\ 
\midrule
\small
Naive & 47.27\% & 89.78\% & 4.28\% & 36.91\% & 8.31\% & High \\
Discrete HMM & 36.20\% & 30.52\% & 14.63\% & 49.32\% & 57.14\% & Medium \\
GraphNN & 56.13\% & 59.60\% & 34.27\% & 66.91\% & 0.00\% & Low  \\
GBM  & 62.54\% & 69.26\% & 0.94\% & 61.56\% & 1.67\% & Low \\
CoTM ($L=0$) & 46.13\% & 37.00\% & 24.68\% & 32.70\% & 0.00\% & High \\
CoTM ($L=1$) & 57.49\% & 50.19\% & 47.27\% & 22.42\% & 0.00\% & High  \\
CoTM ($L=2$) & 68.76\% & 74.24\% & 64.44\% & 26.93\% & 0.00\% & High  \\
Reduced GraphTM & 48.01\% & 66.24\% & 9.93\% & 62.74\% & 0.00\% & High \\
\textbf{Full GraphTM} & 70.69\% & 80.51\% & 48.16\% & 71.91\% & 11.24\% & High \\
\bottomrule
\end{tabular}
\end{table}

The Naive approach is considered the baseline as it simply uses the previous periods regime class as the next prediction (namely the regime identified from the previous 72 hours, so as not to have any forward looking leaks in the data). Since price stability, with low variance and no directional trend tends to dominate currencies most hours of trading, predicting class 0 tends to be the baseline, with occasional stretches of mean reversion that last some time.

Although gradient-boosted models are often effective for tabular prediction tasks, their performance in the present setting was mixed. In particular, the model exhibited difficulty identifying several minority regime classes, suggesting that the available feature representation may not fully capture the temporal dependencies associated with regime transitions. The HMM achieved lower overall accuracy but demonstrated comparatively stronger performance on the rare class 3 observations than several competing methods..  

\begin{figure}[htbp]
	\centering
	\includegraphics[width=0.95\textwidth]{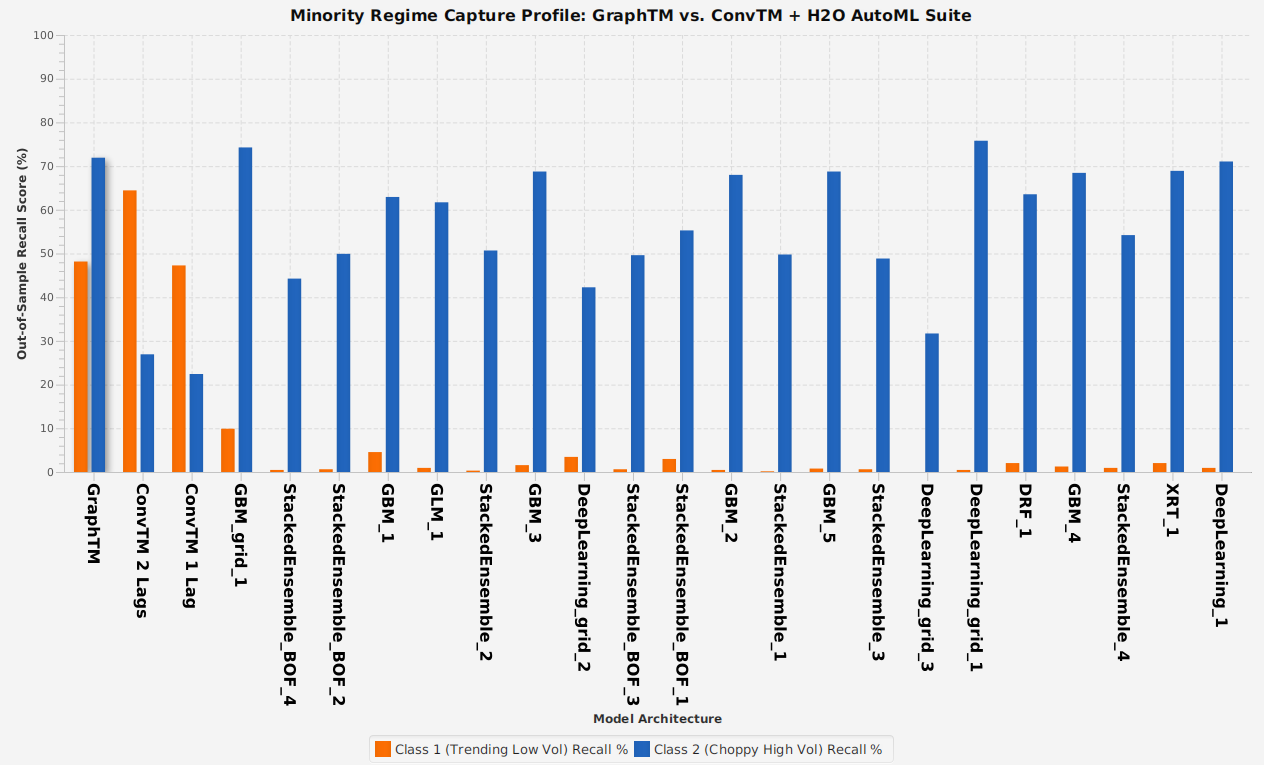}
	\caption{Comparing the Class Regime 1 and 2 OOS accuracy across multiple models including ConvTM with two different lags and AutoML from H2O.}
	\label{fig:allModels}
\end{figure}

In terms of CoTMs performance, the results demonstrate a clear progression of temporal locality. At $L=0$ (utilizing only the instantaneous, concurrent feature state), the model achieves an overall accuracy of 50.13\% but  cannot distinguish or learn any features from  class 0 (0.37\%) and class 3 (0.00\%), while over-indexing on the trend dynamics of class 1 and class 2. Expanding the sliding context window to $L=1$ introduces a singular lag component, which gives a major performance leap to 57.49\% by balancing class 0 prediction (50.19\%). Peak performance for the standard CoTM architecture is achieved at $L=2$, yielding an overall OOS accuracy of 58.76\%.

Finally, the performance divergence between the two topological configurations of the GraphTM yields informative structural insights into the dependencies of currency regimes. While the reduced graph retained the raw \texttt{USD\_NEWS\_SHOCK} node and its direct links to local volatility, it could not identify a high volatile trend. This empirically suggests that macroeconomic shocks cannot be processed as isolated, direct impulses. To decode an high-volatility trend regime shift, the clauses could require the structural intermediary pathways—such as bond market curve repricing and cross-currency risk transmission—to map how an inflation shock diffuses through global liquidity networks. Furthermore, deleting the cross-pair context nodes $\eurusd{}$ and $\audjpy{}$ we see a substantial performance drop in class 1, falling to a mean of 9.93\%. This highlights that internal momentum indicators are fundamentally insufficient for differentiating a stable, quiet macro trend from directionless sideways consolidation (class 0). 

\subsection{Application: Real-time Risk Metrics for Algorithmic Trading}

Standard performance metrics, such as global OOS accuracy or symmetric $F_1$-scores, operate under the implicit assumption of uniform misclassification costs. In algorithmic trading execution systems, this is typically not the case, as there tends to be a higher cost (transactions fees, slippage, trade losses, etc) if employing the wrong strategy for the wrong regime. For example, using a trend following strategy during regime (class 2) could potentially suffer many false directional position entries.  

As an application of our regime detection framework, we formalize a custom risk metric for our GraphTM regime prediction model by constructing an asymmetric cost matrix $C(i, j)$. This cost matrix serves as a prior when trading $\usdjpy{}$.  Let $C(i, j)$ denote the structural penalty incurred by predicting market regime $j \in \{0, 1, 2, 3\}$ when the empirical market reality is regime $i \in \{0, 1, 2, 3\}$

The formulation of the cost matrix $C$ is governed by two axiomatic operational constraints of our trend-following trading engine:
\begin{enumerate}
	\item Activating a high-velocity, momentum-dependent allocation strategy during a highly volatile, mean-reverting cycle ($\mathcal{R}_2$) causes acute capital erosion due to large sequential whipsaws. Thus, a false positive trend prediction given an underlying choppy reality must be heavily penalized: $C(2, 1) = C(2, 3) = \lambda_{\text{whipsaw}}$, where $\lambda_{\text{whipsaw}} \gg 1$.
	\item  Conversely, predicting a structural trend ($\mathcal{R}_1$ or $\mathcal{R}_3$) when the underlying market remains in a low-volatility sideways state ($\mathcal{R}_0$) incurs negligible execution risk. The trading strategy implements downstream execution thresholds and ATR filters that prevent entry orders from filling under low-liquidity conditions, yielding a low baseline penalty: $C(0, 1) = C(0, 3) = \lambda_{\text{filter}} \approx 1$.
\end{enumerate}

For simplicity of exposition, let us assume a trend following strategy on $\usdjpy{}$. Reflecting these asymmetric risk dynamics alongside intermediate operational costs for missed trend opportunities, we formulate an example cost matrix $C$ parameterized as:

\begin{equation}
	C = \begin{pmatrix} 
		C(0,0) & C(0,1) & C(0,2) & C(0,3) \\
		C(1,0) & C(1,1) & C(1,2) & C(1,3) \\
		C(2,0) & C(2,1) & C(2,2) & C(2,3) \\
		C(3,0) & C(3,1) & C(3,2) & C(3,3) 
	\end{pmatrix} = 
	\begin{pmatrix} 
		0 & 1 & 3 & 1 \\
		4 & 0 & 4 & 2 \\
		2 & 10 & 0 & 10 \\
		8 & 2 & 4 & 0 
	\end{pmatrix}
\end{equation}

Here, the cost matrix was constructed to reflect an example of operational objectives of a representative trend-following trading strategy and should therefore be interpreted as application specific rather than universally applicable. The $C(2,1)$ with a weight of 10 represents predicting regime 1 when the actual turned out to be regime 2 (highly choppy and mean reverting). Similarly, when predicting trend but in reality getting regime 0, we also would most likely incur losses. Furthermore, predicting low volatility sideways regime 0, but in reality getting a high variance long trend (regime 3) we have a penalty of 5 due to the opportunity cost of missing the trend.
 
Let $\mathbf{CM}$ be the $4 \times 4$ OOS confusion matrix generated by an empirical model evaluation, where entry $\mathbf{CM}_{i,j}$ represents the total joint frequency of true state $i$ and predicted state $j$. The global \textit{Trading Risk Score} ($\mathcal{R}_{\text{score}}$) is defined as the expectation of the structural cost normalized over the total out-of-sample trading horizon $N$:

\begin{equation}
	\mathcal{R}_{\text{score}} = \frac{1}{N} \sum_{i=0}^{3} \sum_{j=0}^{3} \mathbf{CM}_{i,j} \cdot C(i, j) \quad \text{where} \quad N = \sum_{i=0}^{3} \sum_{j=0}^{3} \mathbf{CM}_{i,j}
\end{equation}

Under this paradigm, the optimal configuration for live deployment is no longer the model that maximizes raw statistical accuracy, but rather the structural function that minimizes expected risk:

\begin{equation}
	f^* = \arg\min_{f \in \mathcal{F}} \mathcal{R}_{\text{score}}(f)
\end{equation}

By transforming the optimization boundary from symmetric error distances to expected capital protection, this objective function effectively penalizes models that suffer from minority-class blindness. We can apply this metric in realtime on a rolling basis to create a risk indicator. Such an indicator can have many applications in trading and risk management, and could even be used to update beliefs in models where the metric is above a certain threshold.  To create our rolling risk indicator, we simply build a OOS confusion matrix over a rolling window of $T$ historical observations up to the most recent observation. As new market observations are received, we update the confusion matrix and the risk score $\mathcal{R}_{\text{score}}$ 

\begin{figure}[htbp]
	\centering
	\includegraphics[width=4in]{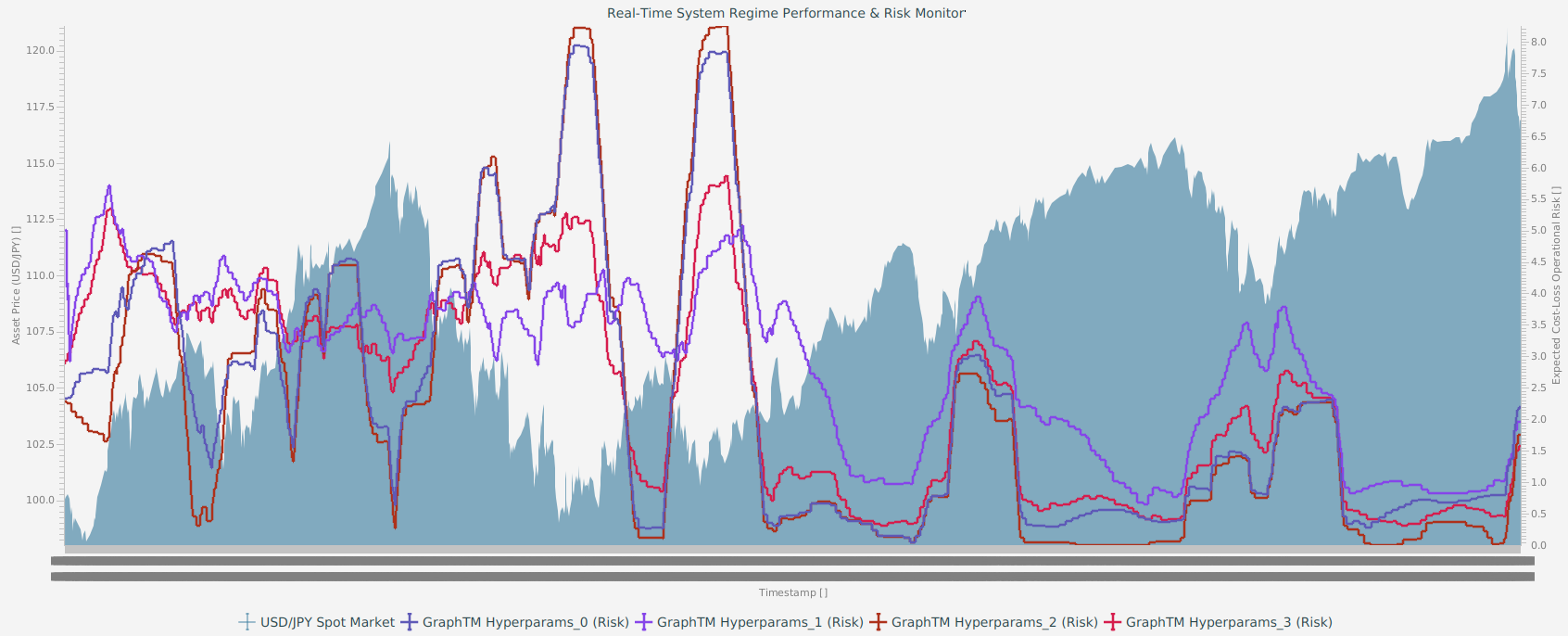}
	\caption{The entire two year period between February 2020 and June 2022. Notice the large amount of deviations in the rolling risk metric during the 2020 period followed by more structural formations with trends.}
	\label{fig:entirePeriod}
\end{figure}

The empirical validity of the proposed dynamic risk framework is evaluated over a macroeconomically turbulent OOS window spanning from the onset of the COVID-19 pandemic in early 2020 through the structural transitions of early 2022. Figure~\ref{fig:entirePeriod} presents the risk metrics of four different GraphTM models with different learning hyperparameter configurations to compare the stability between clause and literal configurations of the nodes and edges. We use a rolling window of 1000 observations. The underlying $\usdjpy{}$ spot market path ($P_t$) is also shown in the background and matched at the timestamp of the risk metric.

\begin{figure}[htbp]
	\centering
	\includegraphics[width=4in]{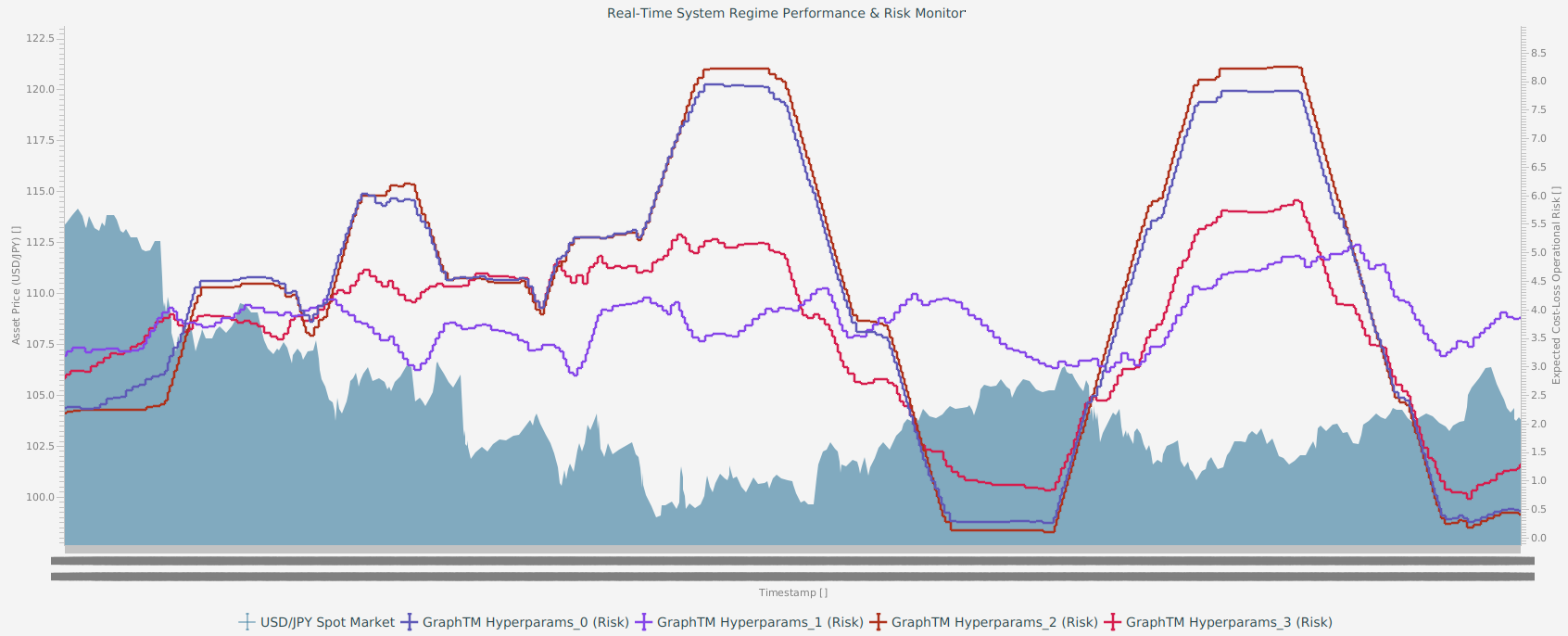}
	\caption{A closeup of the COVID period when regime changes were frequent.}
	\label{fig:covidPeriod}
\end{figure}

During the initial phase of the pandemic in 2020, the asset market exhibited unprecedented volatility, driven by rapid, exogenous macroeconomic shocks and frequent central bank interventions. This systemic instability is directly captured by the risk metric plot, which displays extreme variance and remains consistently elevated toward the final quarter of 2020. This sustained peak indicates a regime of high structural uncertainty where the base classifier experienced elevated prediction entropy, frequently misclassifying highly volatile turning points into costly adjacent states within the confusion matrix. The cost-loss function appropriately penalizes these operational blind spots, establishing a robust quantitative warning signal for strategy degradation during chaotic market regimes.

\begin{figure}[htbp]
	\centering
	\includegraphics[width=4in]{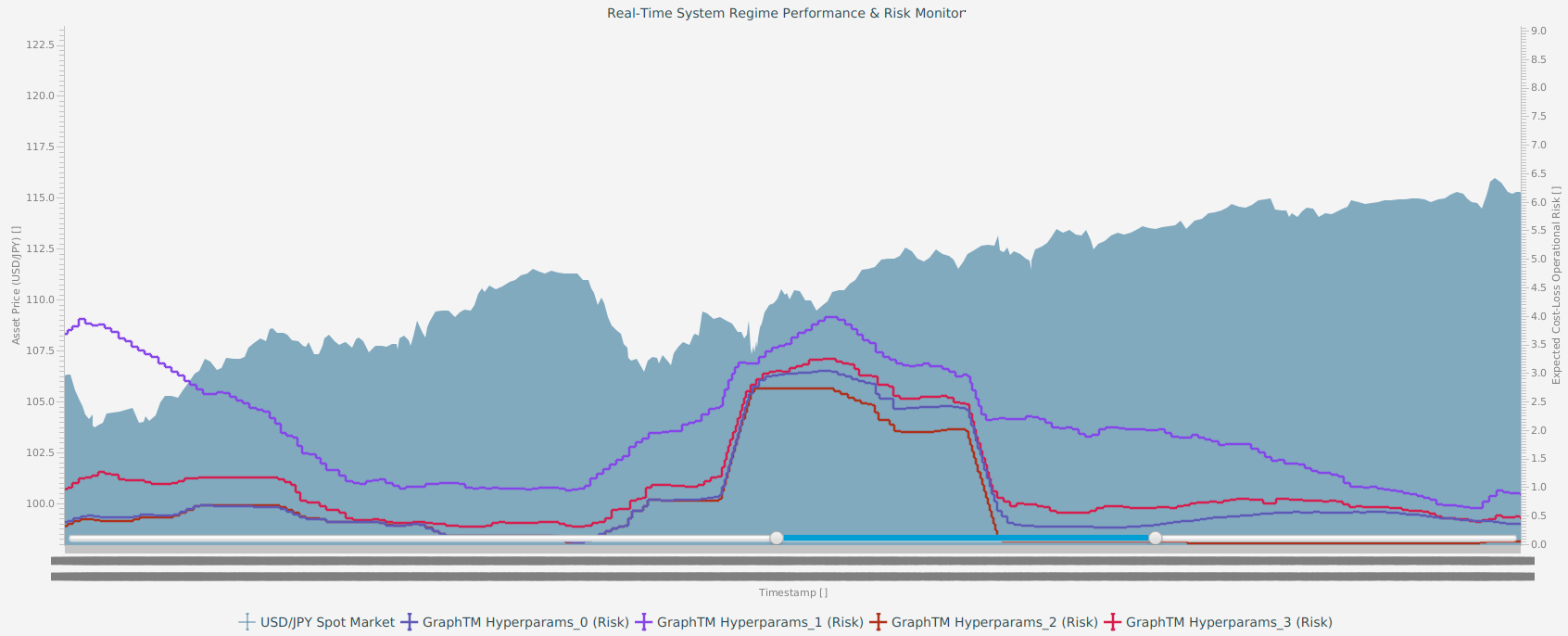}
	\caption{Post COVID period in 2021 and 2022 where regimes were more predictable and thus lower risk profile.}
	\label{fig:period2022}
\end{figure}

Conversely, a distinct structural shift is observed moving into late 2021 and early 2022. As global monetary policies stabilized and the $\usdjpy{}$ pair entered a highly persistent, sustained trend regime, the rolling risk metric decayed sharply, remaining consistently near zero. This prolonged period of near-zero operational risk reflects a highly predictable market macro-cycle, where the spatial-temporal signatures parsed by the GraphTM aligned closely with prevailing market conditions during this period. The minimal overlap in the rolling confusion matrix during this phase could indicate that when the market enters a stable, prolonged regime, the localized cost-loss framework appears to identify the reduction in structural risk, suggesting its utility as a dynamic, real-time capital allocation filter.

\section{Conclusion}
\label{sec:conclusion}

In this paper, we have introduced a novel, topologically grounded framework for macroeconomic foreign exchange regime tracking using the Graph Tsetlin Machine (GraphTM). By mapping decentralized relational cross-market dependencies over hypervectorized directed graphs, our approach departs from both temporal sequence modeling and continuous neural architectures. 

Our empirical evaluations confirm that the proposed graph framework achieves superior predictive capabilities, delivering a mean accuracy competitive to other standard machine learning approaches using the same features. Furthermore, the full macro-topological framework outperformed CoTM architectures, including the CoTM that had a peak baseline at two lags (58.76\%). The CoTM might have experienced clause saturation and performance degradation when expanded beyond two historical lags due to the linear, multiplicative expansion of input literals. Furthermore, our systematic feature ablation study empirically validated the predictive lift provided by global cross-market features ($\eurusd{}$, $\audjpy{}$, $\text{WTI}$ Oil, and US/JP Bond levels. When stripped of these external macro channels, the isolated feature graph suffered an overall performance collapse. 

Future work includes building a similar model across multiple pairs and including even more features into the model. For example, detecting the regime for $\eurusd{}$ and other US dollar pairs, one can use this information in the graph for $\usdjpy{}$, as typically a trending $\eurusd{}$ and $\usdjpy{}$ will imply a mean-reverting $\eurjpy{}$, depending on the scale of the trend. Furthermore, we would also like to extract interpretable rules from the clauses of the model. Can we extract simple rules to determine when a new regime begins? This is the strength of Tsetlin Machines.  In conclusion, the Graph Tsetlin Machine bridges the long-standing gap between structural geometric learning and symbolic explainability, offering quantitative finance and macroeconomic forecasting an exceptionally robust, transparent, and computationally efficient paradigm for navigating multi-modal market regimes.

\defbibnote{preamble}{}

\printbibliography[prenote={preamble}]

\appendix

\section{Node Features}\label{sec:features}

\paragraph{Efficiency Ratio}

The ER at time $t$ over a look-back window of size $n$ is defined as:
\begin{equation}\text{ER}_t = \frac{\left| P_t - P_{t-n} \right|}{\sum_{i=0}^{n-1} \left| P_{t-i} - P_{t-i-1} \right|}\end{equation}
where $P_t$ is the price at the current hour, $P_{t-n}$ is the price at the start of the window.

The resulting value is bounded such that $0 \leq \text{ER}_t \leq 1$. A value of $1.0$ indicates a perfectly efficient, straight-line directional move, while a value approaching $0$ indicates a highly inefficient, mean-reverting, or stagnant market state where price travels a great distance only to return to near its starting point. 

\paragraph{Average True Range}

The Average True Range (ATR), originally introduced by J. Welles Wilder Jr., serves as a non-directional, multi-modal metric to quantify asset volatility by incorporating point-in-time price gaps alongside intraday variance. 

Let $H_t$, $L_t$, and $C_t$ denote the high price, low price, and closing price of the target asset (e.g., $\usdjpy{}$ at chronological time interval $t$, respectively. The \textit{True Range} ($\text{TR}_t$) for a given sample hour $t$ is defined as the maximum absolute distance spanned across three distinct structural price intervals:

\begin{equation}
\text{TR}_t = \max \left\{ (H_t - L_t), \, |H_t - C_{t-1}|, \, |L_t - C_{t-1}| \right\}
\end{equation}

Where:
\begin{itemize}
    \item $(H_t - L_t)$ captures the current period's internal directional trading range.
    \item $|H_t - C_{t-1}|$ measures the absolute distance of an upward price gap relative to the preceding close.
    \item $|L_t - C_{t-1}|$ measures the absolute distance of a downward price gap relative to the preceding close.
\end{itemize}

To smooth out high-frequency micro-noise and capture persistent volatility environments across the temporal network, the individual tracking metrics are smoothed over a lookback window of length $N$. Wilder's standard formulation utilizes a modified exponential moving average (equivalent to an Exponential Moving Average with a smoothing factor $\alpha = 1/N$). 

The \textit{Average True Range} ($\text{ATR}_t$) at time interval $t$ is calculated recursively as follows:

\begin{equation}
\text{ATR}_t = \frac{\text{ATR}_{t-1} \times (N - 1) + \text{TR}_t}{N}
\end{equation}

For the initial seed value of the timeline sequence ($t = N$), the baseline is established via a standard arithmetic mean over the introductory window:

\begin{equation}
\text{ATR}_N = \frac{1}{N} \sum_{i=1}^{N} \text{TR}_i
\end{equation}

In our graph architecture, $N$ is fixed at an hourly scale to match the macro-trajectory adjustments of the network. The resulting $\text{ATR}_t$ values provide the absolute, continuous volatility foundation from which our localized delta layers ($\Delta\text{ATR}_t$) are derived.

\section{Sparse Binary Hypervectors for Financial Graph Structures}\label{sec:hypervectors}

In our framework, nodes carry multivariate continuous features such as normalized volatility and efficiency ratios, while edges are annotated with categorical types representing macroeconomic effects (e.g., interest rate parity or CPI shock), all while message carry clause output information to parent nodes. Sparse Binary Hypervectors (SBH) offer a robust framework for encoding this information due to their composability and inherent noise tolerance. First used in the context of Tsetlin Machines in \cite{halenka2024exploring}, we employ them in our graph framework for computing with economic variables. For more treatment on their computational benefits, we recommend \cite{yang2023cognitivemodelinglearningsparse}.

Represented as high-dimensional vectors $\mathbf{v} \in \{0,1\}^D$ with a small proportion of active bits ($K \ll D$), SBHs ensure that small perturbations in market data do not drastically affect the structural logic of the deep clauses. Furthermore, they enable a symbolic algebra where binding operations preserve the relationship between drivers and targets, while bundling allows the aggregation of multiple market influences into a single, stable representation.

\subsection{Definition and Core Operations}

We define an SBH as a vector where the number of active bits is fixed by a sparsity constant $K$:
\[
||\mathbf{v}||_0 = K, \quad \text{where } \mathbf{v} \in \{0,1\}^D.
\]
In this work, we will take $K$ to be 320 with the full dimension size of 3200, represented efficiently as 50 long 64 bit integers.   

\paragraph{Binding ($\otimes$)}
The \emph{binding} operator $\otimes$ is used to associate a market attribute with a specific context, such as binding a volatility signal to a ``causal'' edge type. Binding is defined as element-wise XOR:
\[
\mathbf{v}_{\text{\tiny bind}} = \mathbf{v}_1 \otimes \mathbf{v}_2 := \mathbf{v}_1 \oplus \mathbf{v}_2.
\]
This operation is invertible ($\mathbf{v}_1 \otimes \mathbf{v}_2 \otimes \mathbf{v}_2 = \mathbf{v}_1$), which is critical for the GraphTM to trace messages back through the economic topography.

\paragraph{Bundling ($\oplus$)}
The \emph{bundling} operator $\oplus$ aggregates a collection of hypervectors—such as multiple macroeconomic drivers affecting the price node—into a single prototype. For each bit position $j \in \{1, \ldots, D\}$, we define the bit-wise sum $c_j = \sum_{i=1}^{M} v_{i,j}$. The bundled vector $\mathbf{v}_{\text{\tiny bundle}}$ is constructed by selecting the top-$K$ positions with the highest counts:
\[
v_{\text{\tiny bundle}, j} =
\begin{cases}
	1 & \text{if } j \in \text{TopK}(c_1, \ldots, c_D) \\
	0 & \text{otherwise}.
\end{cases}
\]
This operation, denoted $\bigoplus_{i=1}^M \mathbf{v}_{i}$, preserves the most common market features while maintaining the required sparsity for logical learning.

\subsection{Embedding Strategies}

To represent continuous and categorical financial data within the GraphTM, we utilize specialized embedding strategies.

\paragraph{Linear Embedding for Market Indicators}
The \emph{linear embedding} maps real-valued market scalars $x \in [a, b]$, such as the log-return of $v_{\text{\tiny UJ}}$ or normalized ATR, into $\mathbf{v}_x \in \{0,1\}^D$. This ensures that similar market states (e.g., two hours of similarly low volatility) map to hypervectors with small Hamming distances. We partition the indicator's historical range $[a, b]$ into $Q$ equally spaced subintervals $\Delta = \frac{b - a}{Q}$. A base vector $\mathbf{v}_{x_0}$ is generated randomly, and each subsequent level is derived via a bit-flipping process:
\[
\mathbf{v}_{x_{i+1}} = \operatorname{FlipBits}(\mathbf{v}_{x_i}, \alpha D + \beta D)
\]
where $\alpha$ is the continuity rate ensuring smooth transitions between adjacent market states, and $\beta$ is the noise rate preventing the model from over-fitting to infinitesimal price fluctuations.

\paragraph{Interval and Categorical Embedding}
To embed discrete attributes, such as specific currency symbols or edge roles (causal vs. correlative), each category $x_q$ is assigned a randomly generated vector:
\[
\phi_{\text{\tiny interval}}(x_q) = \text{RSV}(K)
\]
where $\text{RSV}$ denotes a \textit{Random Sparse Vector}. These vectors are approximately orthogonal, ensuring that the GraphTM treats distinct economic drivers as logically separate entities unless they are explicitly bound or bundled.

To encode information from a node with an edge an adjacent node, as we will do when message passing from one layer to the next, we using binding operations on the edge. To quickly demonstrate, let \( \psi(v) \in \{0,1\}^D \) denote the encoded representation of node \( v \). We construct this using the following composition:
\[
\psi(v) := \phi_{\text{cat}}(\ell_v) \otimes \phi_{\text{lin}}(\mathbf{a}_v) \otimes \phi_{\text{int}}(r_v),
\]
where \( \otimes \) denotes the binding (XOR) operation.

Doing this across the full graph, we simulate attention-style message passing from each vertex \( v \in V \). For each outgoing edge \( e = (v, u) \), we compute a triple binding of:
\begin{itemize}
	\item Source vertex representation \( \psi(v) \),
	\item Edge label and role vector \( \phi_{\text{cat}}(\ell_e) \otimes \phi_{\text{embed}}(e) \),
	\item Target vertex representation \( \psi(u) \).
\end{itemize}

This message is defined as:
\[
\mathbf{m}_{v \rightarrow u} = \psi(v) \otimes \phi_{\text{cat}}(\ell_e) \otimes \phi_{\text{embed}}(e) \otimes \psi(u) \otimes \phi_{\text{embed}}(v),
\]
where the final binding with \( \phi_{\text{embed}}(v) \) ensures role specificity.

All such messages are aggregated using sparse bundling:
\[
\mathbf{v}_G = \bigoplus_{(v,u) \in E} \mathbf{m}_{v \rightarrow u},
\]
where \( \oplus \) is the sparse weighted bundling operator that retains the top \( K \) most frequent bits across messages, maintaining fixed sparsity.

\begin{figure}  
	\centering
	\includegraphics[width=3in]{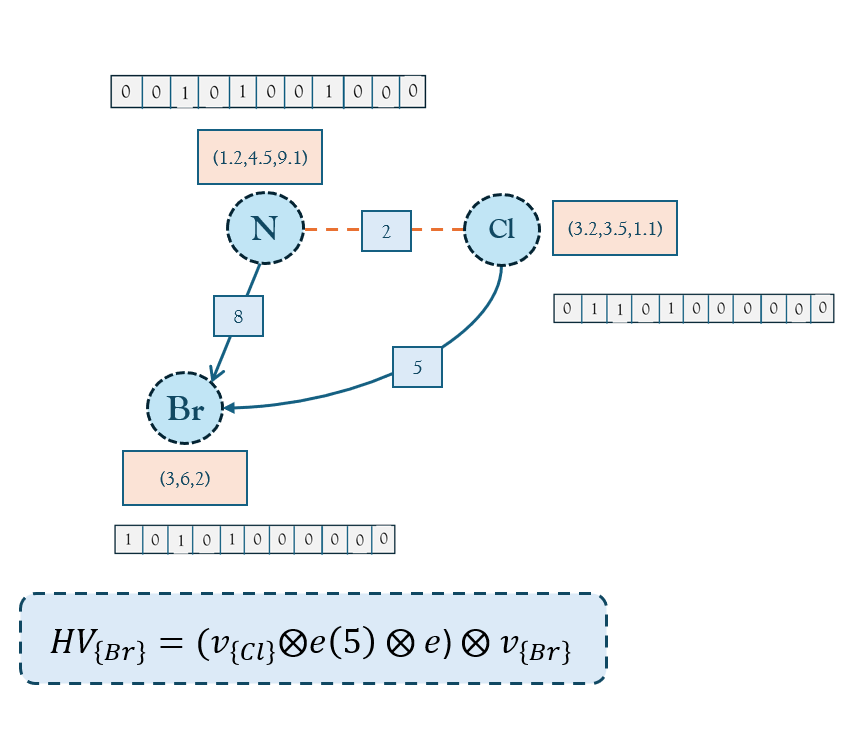}
	\caption{\footnotesize Example of encoding rich information from one node to parent node}
	\label{fig:encoding-nodes}
\end{figure}

Figure \ref{fig:encoding-nodes} illustrates the encoding mechanism from one node to a parent node. As each node label has a categorical role vector associated with it, this is used to bind with the attribute at the node, along with the edge information. Finally, to associate it with the target node, the role vector of the target node is binded to give locational information of the encoding.

\section{Graph Learning Algorithm}\label{sec:algorithm}

A GraphTM is built upon the Convolutional CoTM (see \cite{granmo2019convolutional, glimsdal2021coalesced}), with the following hyperparemeters: depth $d$, number of clauses $n$, specificity $s$, and voting margin $T$. The GraphTM evaluates clauses and nodes in parallel as follows when given an input graph $G$: 
\begin{enumerate}
\item Evaluate each layer-zero clause component for every node in the graph: $C_j^0(v_q), j \in \{1, 2, \ldots, m\}, q \in V$.
\item \textbf{If} $C_j^0(v_q)$ \textbf{then} submit message $\mathbf{M}_j^0 \otimes \mathbf{t}$ to the neighbors of $v_q$ according to $E$, bound to the edge type $\mathbf{t}$.
\item \textbf{For} $i \in (1, 2, \ldots, D)$:
    \begin{enumerate}
        \item Evaluate each layer-$i$ clause component for every node in the graph: $C_j^i(v_q), j \in \{1, 2, \ldots, m\}, q \in V$.
        \item \textbf{If} $C_j^i(v_q)$ \textbf{then} submit message $\mathbf{M}_j^i \otimes \mathbf{t}$ to the neighbors of $v_q$ according to $E$, bound to the edge type $\mathbf{t}$.
    \end{enumerate}
    \item Calculate truth values of full clauses $C_j, j \in \{1, 2, \ldots, m\}$ from the clause components: $C_j = C_j^0 \land C_j^1 \land \cdots \land C_j^D$.
    \item Perform standard Coalesced TM clause voting and update steps using the full clauses and the complete set of properties and messages across the layers.
\end{enumerate}

\subsection{Complete Out-of-Sample Performance results}\label{tab:automl_massive_results}

A complete version of all the out-of-sample performance results from H2O AutoML. Each model considers the same in-sample training data and all are compared across the same highly imbalanced out-of-sample test data.

\begin{longtable}{lcccccr}
	\caption{Comprehensive Out-of-Sample Performance Matrix}   \\
	
	\toprule
	\textbf{Model Archetype} & \textbf{Overall OOS} & \textbf{Class 0} & \textbf{Class 1} & \textbf{Class 2} & \textbf{Class 3} & \textbf{Interpretability} \\
	\midrule
	\endhead 
	
	\toprule
	\multicolumn{7}{c}{\tablename\ \thetable{}} \\
	\midrule
	\textbf{Model Archetype} & \textbf{Overall OOS} & \textbf{Class 0} & \textbf{Class 1} & \textbf{Class 2} & \textbf{Class 3} & \textbf{Interpretability} \\
	\midrule
	\endhead 
	
	\midrule
	\multicolumn{7}{r}{Continued on next page\ldots} \\
	\bottomrule
	\endfoot 
	
	\bottomrule
	\endlastfoot 
	
	\textbf{Full GraphTM} & 70.69\% & 80.51\% & 48.16\% & 71.91\% & 11.24\% & High \\ 
	GLM-11 & 71.99\% & 80.69\% & 0.94\% & 61.71\% & 0.00\% & High \\  
	XRT-3 & 51.77\% & 55.33\% & 2.04\% & 68.91\% & 8.33\% & Low \\ 
	DRF-4 & 52.75\% & 57.07\% & 2.04\% & 63.55\% & 6.67\% & Low \\ 
StackedEnsemble-4 & 61.77\% & 69.09\% & 0.94\% & 54.21\% & 0.00\% & Low \\ 
StackedEnsemble-3 & 68.38\% & 77.64\% & 0.63\% & 48.85\% & 1.67\% & Low \\ 
StackedEnsemble-1 & 52.19\% & 58.02\% & 0.16\% & 49.77\% & 1.67\% & Low \\ 
StackedEnsemble-BOF-4 & 61.32\% & 69.57\% & 0.47\% & 44.26\% & 3.33\% & Low \\ 
StackedEnsemble-2 & 52.67\% & 58.50\% & 0.31\% & 50.69\% & 1.67\% & Low \\ 
StackedEnsemble-BOF-2 & 54.46\% & 60.69\% & 0.63\% & 49.92\% & 3.33\% & Low \\ 
StackedEnsemble-BOF-3 & 55.66\% & 62.18\% & 0.63\% & 49.62\% & 1.67\% & Low \\ 
StackedEnsemble-BOF-1 & 31.72\% & 32.44\% & 2.99\% & 55.28\% & 1.67\% & Low \\
GBM-4 & 43.91\% & 46.02\% & 1.26\% & 68.45\% & 1.67\% & Low \\ 
GBM-5 & 34.33\% & 34.45\% & 0.79\% & 68.76\% & 1.67\% & Low \\ 
GBM-2 & 29.81\% & 29.10\% & 0.47\% & 67.99\% & 1.67\% & Low \\ 
GBM-3 & 39.93\% & 41.15\% & 1.57\% & 68.76\% & 1.67\% & Low \\ 
GBM-53 & 26.57\% & 25.30\% & 4.56\% & 62.94\% & 1.67\% & Low \\ 
GBM-grid-0 & 49.48\% & 53.10\% & 2.04\% & 64.32\% & 0.00\% & Low \\ 
GBM-grid-1 & 33.82\% & 33.48\% & 1.89\% & 71.06\% & 3.33\% & Low \\ 
GBM-grid-2 & 62.54\% & 69.26\% & 0.94\% & 61.56\% & 1.67\% & Low \\ 
GBM-grid-3 & 21.89\% & 18.28\% & 2.67\% & 78.25\% & 1.67\% & Low \\ 
GBM-grid-4 & 52.15\% & 55.98\% & 1.42\% & 68.45\% & 0.00\% & Low \\ 
GBM-grid-5 & 69.13\% & 78.96\% & 0.47\% & 44.72\% & 3.33\% & Low \\ 
GBM-grid-6 & 28.85\% & 27.85\% & 0.00\% & 69.53\% & 0.00\% & Low \\ 
GBM-grid-7 & 16.08\% & 12.53\% & 1.57\% & 66.77\% & 1.67\% & Low \\ 
GBM-grid-8 & 35.80\% & 35.67\% & 2.67\% & 72.44\% & 1.67\% & Low \\ 
GBM-grid-9 & 42.03\% & 43.13\% & 1.26\% & 74.43\% & 3.33\% & Low \\ 
GBM-grid-10 & 27.61\% & 26.35\% & 2.83\% & 65.70\% & 11.67\% & Low \\ 
GBM-grid-11 & 31.65\% & 29.75\% & 9.91\% & 74.27\% & 3.33\% & Low \\ 
GBM-grid-12 & 71.69\% & 80.84\% & 0.94\% & 56.51\% & 0.00\% & Low \\ 
GBM-grid-13 & 19.42\% & 16.52\% & 1.10\% & 67.84\% & 0.00\% & Low \\ 
GBM-grid-14 & 30.32\% & 28.76\% & 1.89\% & 76.11\% & 1.67\% & Low \\ 
GBM-grid-15 & 30.62\% & 29.09\% & 7.07\% & 71.21\% & 3.33\% & Low \\ 
GBM-grid-16 & 53.14\% & 56.97\% & 7.55\% & 64.17\% & 3.33\% & Low \\ 
GBM-grid-17 & 56.13\% & 60.00\% & 5.50\% & 71.82\% & 5.00\% & Low \\ 
GBM-grid-18 & 43.31\% & 45.18\% & 2.67\% & 68.30\% & 0.00\% & Low \\ 
GBM-grid-19 & 48.61\% & 51.01\% & 5.66\% & 70.29\% & 8.33\% & Low \\ 
GBM-grid-20 & 47.02\% & 49.33\% & 5.66\% & 68.30\% & 5.00\% & Low \\ 
GBM-grid-21 & 35.79\% & 39.01\% & 14.94\% & 26.34\% & 11.67\% & Low \\ 
GBM-grid-22 & 31.96\% & 31.11\% & 10.38\% & 64.32\% & 0.00\% & Low \\ 
DeepLearning-2 & 44.93\% & 47.04\% & 0.94\% & 71.06\% & 0.00\% & Low \\ 
DeepLearning-grid-1 & 65.77\% & 72.31\% & 6.60\% & 64.78\% & 0.00\% & Low \\ 
DeepLearning-grid-2 & 72.80\% & 82.23\% & 4.40\% & 52.68\% & 0.00\% & Low \\ 
DeepLearning-grid-3 & 67.00\% & 74.22\% & 0.16\% & 66.77\% & 0.00\% & Low \\ 
DeepLearning-grid-12 & 51.55\% & 56.16\% & 0.31\% & 60.49\% & 0.00\% & Low \\ 
DeepLearning-grid-13 & 52.51\% & 55.30\% & 0.00\% & 80.86\% & 0.00\% & Low \\  
DeepLearning-grid-15 & 19.05\% & 15.32\% & 0.47\% & 75.80\% & 0.00\% & Low \\ 
DeepLearning-grid-16 & 64.88\% & 73.80\% & 3.46\% & 42.27\% & 0.00\% & Low \\ 
DeepLearning-grid-17 & 28.94\% & 25.25\% & 6.92\% & 89.59\% & 0.00\% & Low \\ 
DeepLearning-grid-18 & 9.61\% & 1.64\% & 0.00\% & 98.78\% & 0.00\% & Low \\ 
	\hline
	
\end{longtable}

\end{document}